\documentclass[useAMS,usenatbib,usegraphicx]{mn2e}

\def\simlt{\lower.5ex\hbox{$\; \buildrel < \over \sim \;$}}
\def\simgt{\lower.5ex\hbox{$\; \buildrel > \over \sim \;$}}

\def\beq{\begin{equation}}
\def\eeq{\end{equation}}
\def\del{\delta}
\def\ba{\begin{eqnarray}}
\def\ea{\end{eqnarray}}

\def\c2{c_{200}}

\def\r2{r_{200}}

\def\M2{M_{200}}

\def\zr{z_{\rm rei}}

\title{Studying the Sources of Cosmic Reionization with 21-cm
Fluctuations} \author[Rennan Barkana]{Rennan
Barkana$^{1,2,3}$\thanks{E-mail: barkana@wise.tau.ac.il}\\ $^{1}$
Institute for Cosmic Ray Research, University of Tokyo, Kashiwa
277-8582, Japan \\ $^{2}$ Division of Physics, Mathematics and
Astronomy, California Institute of Technology, Mail Code 130-33,
Pasadena, CA 91125, USA \\ $^{3}$ Guggenheim Fellow; on sabbatical
leave from the School of Physics and Astronomy, Tel Aviv University,
Israel}

\begin{document}

\pagerange{\pageref{firstpage}--\pageref{lastpage}} \pubyear{2008}
\maketitle
\label{firstpage}
\begin{abstract}
We explore the ability of measurements of the 21-cm power spectrum
during reionization to enable the simultaneous reconstruction of the
reionization history and the properties of the ionizing sources. For
various sets of simulated 21-cm observations, we perform maximum
likelihood fits in order to constrain the reionization and galaxy
formation histories. We employ a flexible six-parameter model that
parametrizes the uncertainties in the properties of high-redshift
galaxies. The computational speed needed is attained through the use
of an analytical model that is in reasonable agreement with numerical
simulations of reionization. We find that one-year observations with
the MWA should measure the cosmic ionized fraction to $\sim 1\%$
accuracy at the very end of reionization, and a few percent accuracy
around the mid-point of reionization. The mean halo mass of the
ionizing sources should be measurable to $10\%$ accuracy when
reionization is 2/3 of the way through, and to $20\%$ accuracy
throughout the central stage of reionization, if this mass is anywhere
in the range 1/3 to 100 billion solar masses.
\end{abstract}

\begin{keywords}
galaxies:high-redshift -- cosmology:theory -- galaxies:formation
\end{keywords}

\section{Introduction}\label{intro}

The earliest generations of galaxies are thought to have heated and
reionized the universe. Ly$\alpha$ absorption shows that the IGM has
been a hot plasma at least since $z \sim 6.5$ \citep{fan06}, while the
five-year WMAP measurements of the large-angle polarization of the
cosmic microwave background imply that the universe was significantly
ionized within the redshift range $z \sim 8$--14 \citep{wmapRei}. The
same WMAP measurements are also consistent with the $\Lambda$CDM model
and, together with distance measurements from supernovae and baryon
acoustic oscillations from galaxy surveys, have helped to determine
rather accurately the standard cosmological parameters of this model
\citep{wmap}; we thus assume the $\Lambda$CDM model with density
parameters $\Omega_m=0.28$ (dark matter plus baryons),
$\Omega_\Lambda=0.72$ (cosmological constant), and $\Omega_b=0.046$
(baryons), together with $h=0.7$ (dimensionless Hubble constant),
$n=0.96$ (power spectrum index), and $\sigma_8=0.82$ (power spectrum
normalization).

Observational study of reionization promises to teach us a great deal
about the early generations of galaxies that formed when the universe
was between $\sim 300$ and 800 Myr in age. An important feature of
reionization is that it indirectly probes whatever are the dominant
sources of ionizing radiation, even if these are otherwise
unobservable; e.g., the universe may have been ionized by large
numbers of very small galaxies that are too faint to be detected
individually (or even by unexpected sources such as miniquasars or
decaying dark matter). The overall timing of reionization versus
redshift mainly constrains the overall cosmic efficiency of ionizing
photon production. In comparison, a detailed picture of reionization
as it happens can teach us a great deal about the sources that
produced this cosmic phase transition, particularly in the case of the
most natural source, stars in galaxies.

A key point is that the spatial distribution of ionized bubbles is
determined by clustered groups of galaxies and not by individual
galaxies. At such early times galaxies were strongly clustered even on
rather large scales (tens of comoving Mpc), and these scales therefore
dominate the structure of reionization \citep{BLflucts}. Overdense
regions fully reionize first because the number of ionizing sources in
these regions is increased very strongly \citep{BLflucts}. The
large-scale topology of reionization is therefore inside out, with
underdense voids reionizing only at the very end of reionization, with
the help of extra ionizing photons coming in from their surroundings.
This picture has been confirmed and quantified more precisely in
detailed analytical models that account for large-scale variations in
the abundance of galaxies \citep{fzh04}, and in a number of
large-scale numerical simulations of reionization
\citep[e.g.,][]{zahn, iliev, santos}.

Cosmic reionization represents an extreme challenge for numerical
simulations, because of the enormous range of spatial scales involved
\citep{BLflucts}. On the one hand, individual galaxies as small as a
comoving kpc may contribute to the early stages of reionization, while
ionizing photons may travel as much as 100 Mpc before being absorbed,
near the end of reionization (or earlier if X-rays make a significant
contribution). Even smaller scales must be resolved in order to probe
the dense gas clumps that likely determine the ionizing mean free path
and the recombination rate, while star formation and stellar feedback
are still further out of reach. While simulations are improving and
may soon include hydrodynamics (in addition to the current N-body plus
radiative transfer codes), some have tried to bridge the gap of scales
by developing fast semi-numerical methods that make substantial use of
approximate analytical models \citep{zahn, mesinger}.

The most promising probe of the cosmic reionization history is to use
new low-frequency radio telescope arrays to detect emission in the
redshifted 21-cm line corresponding to the hyperfine transition of
atomic hydrogen. 21-cm cosmology is potentially also a great source of
fundamental cosmological information, especially if observations reach
small scales and high redshifts. The 21-cm fluctuations can in
principle be measured down to the smallest scales where the baryon
pressure suppresses gas fluctuations, while the cosmic microwave
background (CMB) anisotropies are damped on much larger scales
(through Silk damping and the finite width of the surface of last
scattering). Since the 21-cm technique is also three-dimensional
(while the CMB yields a single sky map), there is a much larger
potential number of independent modes probed by the 21-cm signal,
which could help to detect primordial non-Gaussianity and test
inflation \citep{zald}. However, ionization fluctuations dominate the
21-cm fluctuations during the epoch of reionization, and thus the
first generation of 21-cm experiments are expected to bring new
discoveries related to the reionization history rather than the
fundamental cosmological parameters
\citep[e.g.,][]{mcquinn06}. Furthermore, since the observational noise
will be too large to produce 21-cm maps, the fluctuations must be
measured statistically, and the power spectrum is both the most
natural and the highest signal-to-noise statistic. Upcoming
experiments include the Murchison Widefield Array
(MWA)\footnote{http://www.haystack.mit.edu/ast/arrays/mwa/} and the
Low Frequency Array (LOFAR)\footnote{http://www.lofar.org/}.

While theorists and numerical simulators have begun to elucidate the
relation between the properties of the ionizing sources and the 21-cm
power spectrum, a key question has not been addressed thus far:
assuming the upcoming observations measure the 21-cm power spectrum as
expected, how is the power spectrum to be inverted into a
determination of the properties of the sources? Such an inversion
problem is usually solved by a maximum likelihood (or $\chi^2$)
procedure whereby a model is fit to the observed power spectrum in
order to determine the best-fit parameters and their uncertainties. In
order to explore maximum likelihood fitting of simulated observations,
a flexible model is needed that can quickly yield the 21-cm power
spectrum predicted for given parameters of the ionizing galaxies. It
is important to have a flexible model that does not presume that we
can theoretically predict the properties of the ionizing galaxies,
which depend on many complex feedback processes. This
``reionize-fast'' code would essentially play the same role that
CMBFAST \citep{cmbfast} did for analyses of measurements of the CMB
angular power spectrum.  Ultimately, this type of code will most
likely be developed from an analytical model that includes as much of
the detailed physics as possible and is also partly tuned to fit more
accurately the results of numerical simulations, analogous to the way
that the formula of \citet{Sheth} for the halo mass function was
developed from the original model of \citet{PS}.

In this paper we employ the model from \citet{b07} in which we solved
for the correlated two-point distribution of density and ionization
based on the one-point model of \citet{fzh04}. This is currently the
most realistic fully analytical model of the 21-cm power spectrum. In
the next section we briefly review the model before comparing its
predictions to 21-cm power spectra from several numerical simulations
of reionization. In the following section we use the model to
summarize which galaxy parameters affect the 21-cm power spectrum, and
then calculate the expected uncertainties from maximum likelihood fits
to simulated sets of observed power spectra.

\section{The Model: Setup and Tests}

\subsection{The analytical model}

\label{s:model}

Analytical approaches to galaxy formation and reionization are based
on the mathematical problem of random walks with barriers. The basic
approach is that of \citet{bc91}, who rederived and extended the halo
formation model of \citet{PS}. The idea is to consider the smoothed
density in a region around a fixed point in space.  We begin by
averaging over a large (comoving) scale $R$, or, equivalently, by
including only small comoving wavenumbers $k$. We then lower $R$,
generating a random walk as power on smaller scales (higher $k$
values) is added, until we find the first (i.e., largest) scale for
which the averaged overdensity is high enough to reach a particular
milestone. The needed overdensity is termed the barrier, and the goal
is then to find the distribution of points at which the random walk
first crosses the barrier. For the halo mass function, spherical
collapse yields a constant barrier (i.e., the required initial
overdensity, linearly extrapolated to the present, is independent of
halo mass or scale $R$).

\citet{fzh04} showed that the condition of having enough ionizing
sources to fully ionize a region corresponds approximately to a linear
barrier, and then used the statistics of a random walk with a linear
barrier to predict the H~II bubble size distribution during the
reionization epoch. In \citet{b07} we found an accurate analytical
solution for the corresponding two-point problem of two correlated
random walks with linear barriers, using the two-step approximation
which \citet{sb} had applied to the two-point constant barrier
problem. Finding the joint probability distribution of the density and
ionization state of two points allows the calculation of the 21-cm
correlation function or power spectrum \citep{b07}.

Following \citet{fzh04}, the appropriate barrier for reionization is
found by setting the ionized fraction in a region $\zeta F_{\rm coll}$
equal to unity, where $F_{\rm coll}$ is the collapse fraction (i.e.,
the gas fraction in galactic halos) and $\zeta$ is the overall
efficiency factor, which is the number of ionizing photons that escape
from galactic halos per hydrogen atom (or ion) contained in these
halos. This simple version of the model remains approximately valid
even with recombinations if the effective $\zeta$ is divided by one
plus the number of recombinations per hydrogen atom in the IGM,
assuming this factor is roughly uniform. In order to find $F_{\rm
coll}$, a good starting point is the formula of \citet{Sheth}, which
accurately fits the cosmic mean halo abundance in
simulations. However, an exact analytical generalization is not known
for the biased $F_{\rm coll}$ in regions of various mean density
fluctuation $\del$.

\citet{BLflucts} suggested a hybrid prescription that adjusts the
abundance in various regions based on the extended Press-Schechter
formula \citep{bc91}, and showed that it fits a broad range of
simulation results. In general, we denote by $f(\del_c(z),S)\, dS$ the
mass fraction contained at $z$ within halos with mass in the range
corresponding to variance $S$ to $S+d S$, where $\del_c(z)$ is the
critical density for halo collapse at $z$. Then the biased mass
function in a region of size $R$ (corresponding to density variance
$S_R$) and mean density fluctuation $\del$ is \citep{BLflucts} \ba
\lefteqn{f_{\rm bias}(\del_c(z),\delta,R,S) = \frac{f_{\rm
ST}(\del_c(z),S)} {f_{\rm PS}(\del_c(z),S)} \ f_{\rm PS}
(\del_c(z)-\delta,S-S_R) \ ,} \nonumber \\ && \label{eq:bias} \ea
where $f_{\rm PS}$ and $f_{\rm ST}$ are, respectively, the
Press-Schechter and Sheth-Tormen halo mass functions. The value of
$F_{\rm coll}(\del_c(z),\delta,R,S)$ is the integral of $f_{\rm bias}$
over $S$, from 0 up to the value $S_{\rm min}$ that corresponds to the
minimum halo mass $M_{\rm min}$ or circular velocity $V_{\rm
c}=\sqrt{G M_{\rm min}/R_{\rm vir}}$ (where $R_{\rm vir}$ is the
virial radius of a halo of mass $M_{\rm min}$ at $z$). We then
numerically find the value of $\delta$ that gives $\zeta F_{\rm
coll}=1$ at $S=0$ and its derivative with respect to $S$, yielding the
linear approximation to the barrier: $\delta(S) \approx \nu + \mu
S$. Note that \citet{b07} and \citet{diffPDF} used an approximation in
which effectively each factor on the right-hand side of
equation~(\ref{eq:bias}) was integrated separately over $S$, yielding
a simple analytical formula for the effective linear barrier. Here we
solve numerically for the barrier using the exact formulas (though the
difference in the final results is small).

By the reionization epoch, there are expected to be sufficient
radiation backgrounds of X-rays and of Ly$\alpha$ photons so that the
cosmic gas has been heated to well above the cosmic microwave
background temperature and the 21-cm level occupations have come into
equilibrium with the gas temperature \citep{Madau}. In this case, the
observed 21-cm brightness temperature relative to the CMB is
independent of the spin temperature and, for our assumed cosmological
parameters, is given by \citep{Madau} \beq T_b = \tilde{T}_b(z) \Psi;\
\ \ \ \tilde{T}_b(z) = 25 \sqrt{\frac{1+z} {8}}\, {\rm mK}\ ,
\label{eq:Tb}\eeq with $\Psi = x^n [1+D(z) \del]$, where $x^n$ is the
neutral hydrogen fraction and the linear overdensity at $z$ is the
growth factor $D(z)$ times $\del$ (which is the density linearly
extrapolated to redshift 0). Under these conditions, the 21-cm power
spectrum is thus $P_{21} = \tilde{T}_b^2 P_\Psi$, and thus a model of
the relation between the density and the ionization is all that is
needed for calculating the 21-cm power spectrum.

The analytical model thus consists of the following: For a given
efficiency $\zeta$ and minimum halo circular velocity $V_{\rm c}$ at
redshift $z$, find the corresponding linear barrier coefficients $\nu$
and $\mu$, calculate the 21-cm correlation function as a function of
separation $d$ (where at each $d$ we numerically integrate
equation~(49) of \citet{b07}), and then Fourier transform to find the
power spectrum at the desired values of $k$. Even with the analytical
model, this procedure is too slow to apply directly in the $\chi^2$
fitting, but the power spectrum can be interpolated from a large
precomputed table as a function of the three variables $\zeta$,
$V_{\rm c}$, and $z$. Note that our assumption of a fixed $\zeta$ (at
a given $z$) for all halos above the minimum $V_{\rm c}$ is not as
strong a restriction as it may appear. Since the halo mass function
declines rapidly with mass at the high redshifts of the reionization
era, once $V_{\rm c}$ is fixed, most of the ionizing sources are close
in mass (i.e., within a factor of a few) to the minimum mass. Thus,
even if in the real universe $\zeta$ varies with mass at a given
redshift, it is unlikely that the total ionized volume will receive
large contributions from a wide range of halo masses.

With the basic setup just described, we are free to apply any values
of $\zeta$ and $V_{\rm c}$ at various redshifts where the power
spectrum can be observed. The simplest model we use is thus a
two-parameter model where $\zeta$ and $V_{\rm c}$ are both assumed to
be constant with redshift. However, complex, time-variable feedbacks
are likely to be operating during reionization, such as X-ray and UV
photo-heating, supernovae and stellar winds, metal enrichment (and the
consequent changes in gas cooling and stellar populations), feedback
from mini-quasars, and radiative feedbacks that affect $H_2$ formation
and destruction. Many of these feedbacks involve scales that are far
too small for direct numerical simulation, certainly within a
cosmological context, so instead of trying to use particular models we
prefer to parametrize our ignorance using additional free
parameters. The third parameter that we add is a coefficient that
gives $V_{\rm c}$ a linear dependence on $z$, and the fourth allows a
linear redshift-dependence in $\zeta$. Similarly, a fifth and sixth
parameter allow a quadratic redshift-dependence in $\zeta$ and $V_{\rm
c}$, thus permitting these parameters to vary more flexibly with
redshift (including a slope that may even change in sign during
reionization). Our main goal is to see whether the 21-cm power
spectrum can help determine both the reionization history and key
properties of the ionizing sources, even if we allow for such flexible
models of the ionizing sources with six free parameters that are not
restricted based on specific models of feedback.

\subsection{Comparison with numerical simulations}

Numerical simulations of reionization are a rapidly developing
field. Current simulations are based on purely gravitational N-body
codes that are used to locate and weigh forming halos as a function of
time. Radiative transfer codes are then used to find the reionization
topology due to ionizing photons coming from the source halos. Thus,
simulations offer the potential advantages of fully realistic source
halo distributions and accurate radiative transfer. Resources, though,
are still stretched when attempts are made to resolve the smallest
source halos while having sufficiently large boxes for tracking
ionizing photons with the longest mean free paths. Also, while
prospects are good for also including hydrodynamics, it seems that
astrophysics for the foreseeable future must be included
schematically, as in an analytical model. The important aspects of
astrophysics that are inserted by hand include at least the star
formation rate within each halo, properties of the stellar
populations, supernova feedback (including suppression of star
formation, metal enrichment, and dust formation), photo-heating
feedback, and the escape of ionizing photons from each galaxy.

\begin{figure}
\includegraphics[width=84mm]{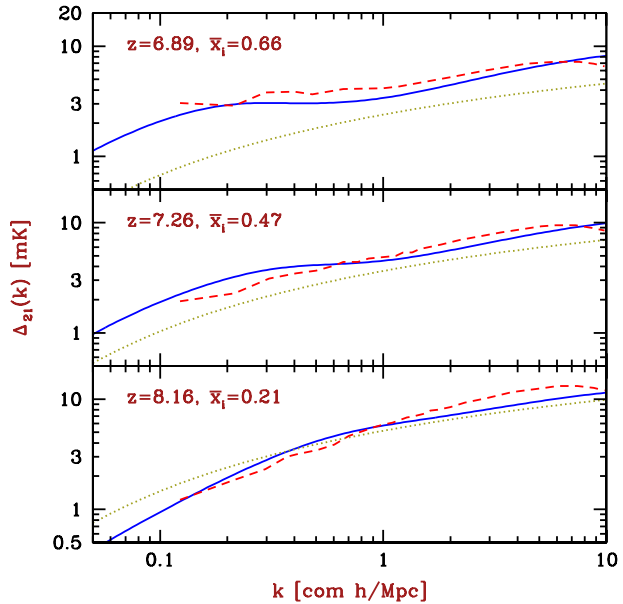}
\caption{The analytical model: Test \#1. We compare our predictions
for the 21-cm power spectrum using the \citet{b07} model (solid
curves) to those from the simulation of \citet{zahn} (dashed
curves). The results are shown at several different redshifts, as
indicated in each panel. At each redshift $z$ we adjust the value of
the efficiency in our model in order to match the cosmic mean
mass-weighted ionized fraction $\bar{x}^i$ from the simulation. Also
shown (dotted curve) is the 21-cm power spectrum for a uniformly
ionized universe at the same $\bar{x}^i$.}
\label{f:test1}
\end{figure}

\begin{figure}
\includegraphics[width=84mm]{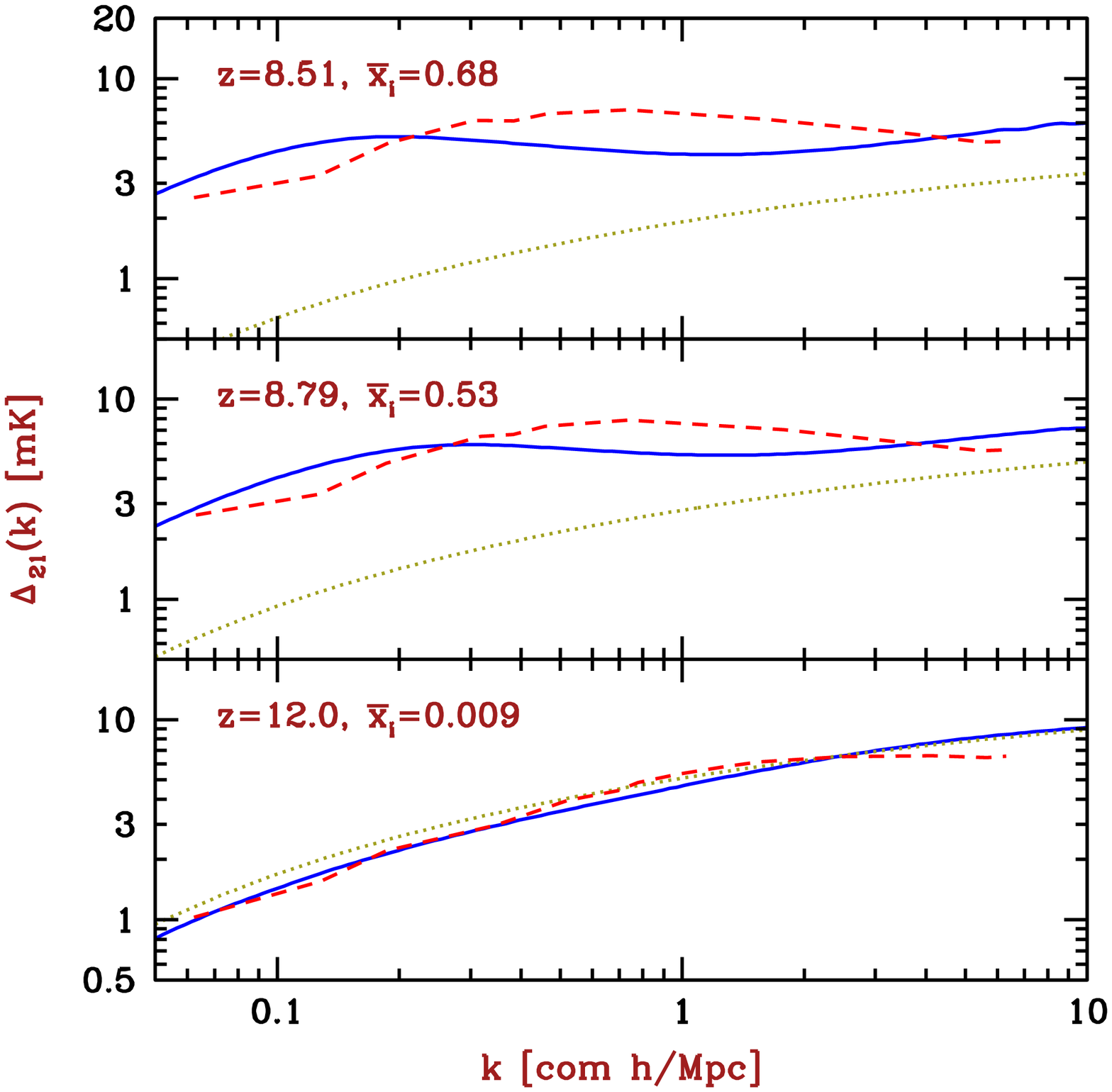}
\caption{The analytical model: Test \#2. Same as Figure~\ref{f:test1}
except that we compare with $P(k)$ from the simulation f250 of
\citet{iliev}.}
\label{f:test2}
\end{figure}

\begin{figure}
\includegraphics[width=84mm]{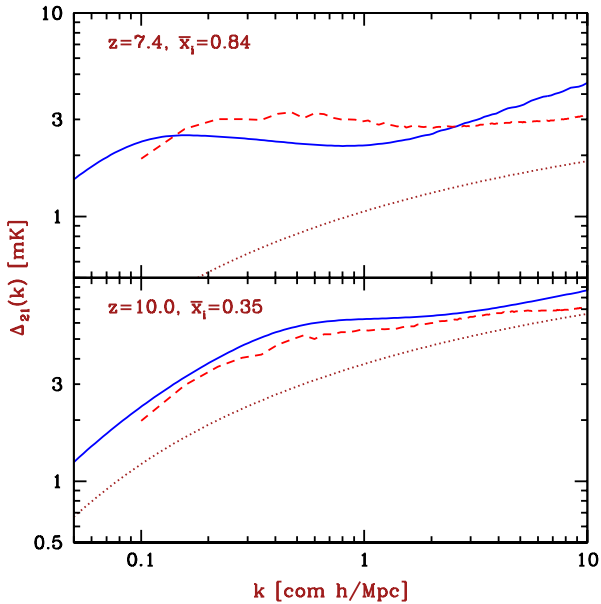}
\caption{The analytical model: Test \#3. Same as Figure~\ref{f:test1}
except that we compare with $P(k)$ from the simulation of
\citet{santos}.}
\label{f:test3}
\end{figure}

Since the analytical model we use is limited in using spherical
statistics as a simple approximation for radiative transfer, it is
useful to compare it to results of numerical simulations. We compare
our 21-cm power spectrum predictions based on \citet{b07} to those
measured in numerical simulations of \citet{zahn}\footnote{Note that a
comparison to these simulations was also shown in Figure~4 of
\citet{b07}, but there a preprint of \citet{zahn} was used for results
which changed in the final published version, and also there the
mass-weighted ionized fraction from the model was incorrectly matched
to the volume-weighted one from the simulation.}, \citet{iliev} and
\citet{santos} in Figures~\ref{f:test1}, \ref{f:test2} and
\ref{f:test3}, respectively. For comparison, the figures also show the
shape of the 21-cm power spectrum if it arose purely from density
fluctuations; the normalization of these curves corresponds to a
uniformly ionizing universe (see also the next subsection). The
figures show the brightness temperature fluctuation \beq
\Delta_{21}(k) \equiv \tilde{T}_b(z)\, \sqrt{k^3 P(k)/(2 \pi^2)}\
. \label{e:d21} \eeq

The simulations are all in reasonable agreement with the analytical
model. The agreement is especially good with \citet{zahn}, where the
typical error in $\Delta_{21}$ is $\sim 10\%$ although it ranges
up to $\sim 25\%$. The agreement with \citet{iliev} is good at $z=12$,
when density fluctuations are completely dominant, but later during
reionization a $25\%$ difference is typical, with the simulation
curves showing a somewhat different shape that includes a decrease
with $k$ at $k \ga 1h$/Mpc. There is good agreement (typically $\sim
10\%$) with \citet{santos} at $z=10$, but at the lower redshift in
Figure~\ref{f:test3} the simulated $\Delta_{21}$ is flat over a
wider range of $k$ than the theoretical one is, and they typically
differ by $\sim 30\%$.

In the comparison, in order to try to match the assumptions in the
simulations as closely as possible, in the first two comparisons we
assumed in the model that the ionizing emission rate from each halo is
proportional to its mass. In test \#3, we instead assumed that the
emission rate is proportional to the gas infall rate into each halo;
this is a more natural assumption within the context of the analytical
model, and we use it in all of our model calculations below. In any
case, the difference between these two assumptions has a minor effect
on the 21-cm power spectrum (for a fixed value of $\bar{x}^i$). We
note that while the analytical model accounts for the restriction of
$x^i$ to a value of 0 or 1, and includes a complex dependence of $x^i$
on $\del$, it neglects the non-linear growth of $\del$. The latter
becomes important only on smaller scales than those accessible to the
first-generation 21-cm experiments. For example, $k \sim 1 h/Mpc$
corresponds to a scale $R \sim 9$ comoving Mpc, which at redshift 8
has a root-mean-square fluctuation of 0.14 (on an observed angular
scale of $3^\prime$). Also, the model does not include the small gas
fraction in leftover neutral clumps within the ionized bubbles, which
is important towards the end of reionization. The simulations do
include some of the leftover neutral gas, but the limited resolution
limits the ability to accurately track these small-scale, non-linear
clumps.

From these comparisons we can conclude that the analytical model
generally captures the evolution of the 21-cm power spectrum during
reionization as seen in the simulations. It is difficult to make a
more quantitative assessment, since the varying results in the
comparison to the different simulations suggest that the simulations
may still disagree among themselves. For further progress, these
simulations must be demonstrated to have numerically converged both
individually and collectively. Individual convergence would mean
showing for each simulation code that the 21-cm power spectrum (over
the relevant scales) is independent of the simulated box size (the
current sizes produce some large-scale fluctuations in $\Delta_{21}$
that are obvious in the figures), the N-body resolution (which affects
halo formation and structure), and the radiative grid
resolution. Collective convergence would mean showing that the various
radiative transfer codes all yield the same power spectrum when
assuming the same initial conditions and halo astrophysics.

Once numerical convergence is demonstrated, it will become possible to
run a suite of simulations in order to test the analytical model more
precisely, and perhaps to develop an improved analytical model based
on fitting the numerical simulation results. Such a model can then be
used in place of the analytical model that we use here, in order to
make more accurate predictions of the 21-cm power spectrum and
eventually to fit the real 21-cm data.

\section{Results}

\subsection{Parameter dependence}

In order to understand what can be learned from observing the 21-cm
power spectrum, a key question is which variables mainly determine the
21-cm power spectrum during the reionization epoch. This question has
been studied in many papers both analytically and from simulations, in
particular by \citet{mcquinn07} and \citet{lidz08}. In this subsection
we use the \citet{b07} model to briefly illustrate the expected power
spectrum evolution for various assumptions regarding the ionizing
source galaxies.

We first note some general properties of the model (see also
\citet{fzh04}). If the universe were uniformly ionized, then we would
have $P_\Psi = \left( \bar{x}^n \right)^2 P_\delta$, where $\bar{x}^n$
is the neutral hydrogen fraction. Fortunately for the observers, as
shown in Figures~\ref{f:test1} through \ref{f:test3}, the power
spectrum is significantly higher in the standard picture where
reionization is caused by stellar radiation, resulting in a
swiss-cheese division of the IGM into ionized bubbles and neutral
regions. In the model we use, on small scales (i.e., much smaller than
the H~II bubbles), $P_\Psi \approx \bar{x}^n P_\delta$, different from
the uniformly ionized case since the average value $\langle (x^n)^2
\rangle = \langle x^n \rangle$ when $x^n$ can only take on the values
0 and 1. On large scales, if we assume linear fluctuations where the
ionized sources have a mean (Lagrangian) bias of $b>0$, so that the
ionized fraction fluctuations are $\del_{x^i} = b \del$, then $P_\Psi
= \left[ 1-\bar{x}^i (1+b) \right]^2 P_\delta$. In regions of high
density, there are potentially more hydrogen atoms but the neutral
fraction is lower, giving rise to an anticorrelation between density
and 21-cm emission when $\bar{x}^i>1/(1+b)$. This transition typically
occurs early in reionization $(\bar{x}^i \sim 0.1-0.15)$ if $b \sim
5-10$ for the ionizing sources.

Thus, $P_{21} \propto P_\del$ on both large and small scales within
the model, but with different normalizations. A transition between
these two regimes occurs on scales of order the bubble size, where the
power spectrum is often rather flat (in terms of $k^3 P_{21}(k)$
varying slowly with $k$). Early on, when $\bar{x}^i \ll 1$, $P_\Psi
\approx P_\del$ on all scales. As reionization proceeds, the bias of
the ionizing sources raises the large-scale 21-cm power spectrum, with
the characteristic bubble (and transition) scale growing with time.
Near the end of reionization, the large-scale regime becomes
irrelevant (as the bubble size diverges within the model), and the
entire power spectrum drops as most of the hydrogen is ionized and no
longer contributes to 21-cm emission. 

\begin{figure}
\includegraphics[width=84mm]{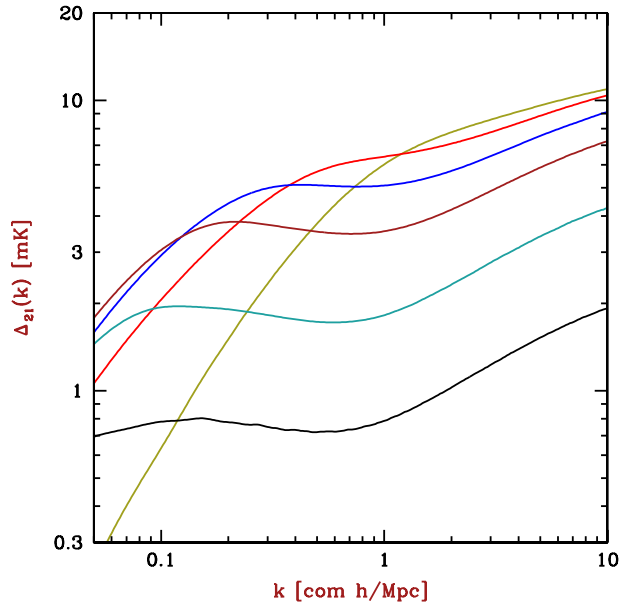}
\caption{Evolution of 21-cm power spectrum throughout reionization,
for a model that sets $\bar{x}^i=98\%$ at $z=6.5$ with minimum $V_{\rm
c} = 35$ km/s. We consider $\bar{x}^i=10\%$, $30\%$, $50\%$, $70\%$,
$90\%$, and $98\%$ (from top to bottom at large $k$).}
\label{f:th35}
\end{figure}

Figure~\ref{f:th35} illustrates the evolution of the 21-cm power
spectrum in a model with a constant (i.e., redshift-independent)
minimum $V_{\rm c} = 35$ km/s for galactic halos, and a constant
efficiency set to put the end-stages of reionization (i.e.,
$\bar{x}^i=98\%$) at $z=6.5$. The Figure shows the predicted power
spectrum at $\bar{x}^i=10\%$ ($z=10.5$), $30\%$ ($z=8.7$), $50\%$
($z=7.8$), $70\%$ ($z=7.1$), $90\%$ ($z=6.7$), and $98\%$
($z=6.5$). We express the result in terms of the fluctuation level in
units of brightness temperature [eq.~(\ref{e:d21})]. We note that for
this model with $V_{\rm c} = 35$ km/s, the overall efficiency is
$\zeta=31$, and at the midpoint of reionization ($\bar{x}^i=50\%$) the
minimum galactic halo mass is $1 \times 10^9 M_\odot$ while the mean
halo mass (weighted by ionization intensity) is $3 \times 10^9
M_\odot$.

\begin{figure}
\includegraphics[width=84mm]{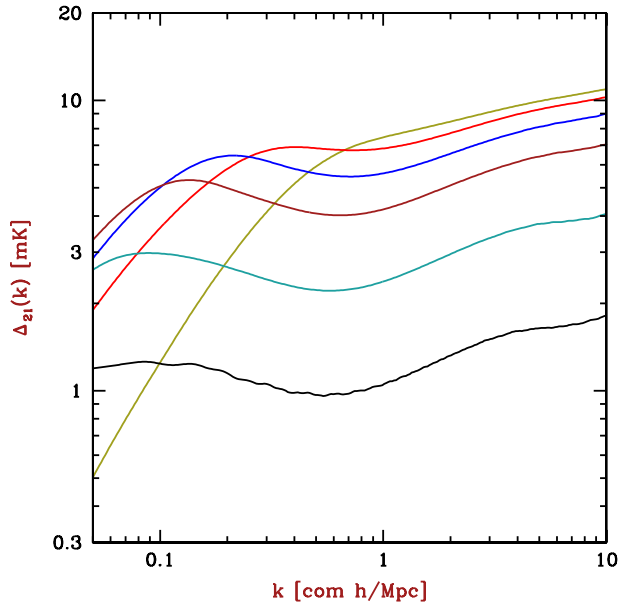}
\caption{Same as Figure~\ref{f:th35} except we set $V_{\rm c} = 100$
km/s.}
\label{f:th100}
\end{figure}

Figure~\ref{f:th100} shows the same but with $V_{\rm c} = 100$ km/s,
i.e, a minimum halo mass higher by a factor of 23. In this case, the
same six values of $\bar{x}^i$ correspond to redshifts $z=9.0$, 7.8,
7.3, 6.9, 6.6, and 6.5. In this model the required efficiency is
$\zeta=195$, and at the midpoint of reionization ($\bar{x}^i=50\%$)
the minimum galactic halo mass is $3 \times 10^{10} M_\odot$ while the
mean halo mass (weighted by ionization intensity) is $5 \times 10^{10}
M_\odot$.

Comparing Figures~\ref{f:th35} and \ref{f:th100} we note that in both
cases the power spectrum evolution is broadly similar when considered
at the same values of $\bar{x}^i$, yet there are important differences
that allow the power spectrum measurements to probe the characteristic
halo mass of the ionizing sources. In particular, more massive halos
are more highly biased, and thus produce a higher small-$k$ ``bump''
during the central stages of reionization. In addition, since more
massive halos are more rare and correspond to the Gaussian tail of
large (positive) density fluctuations, their abundance changes rapidly
with time as density fluctuations grow, thus making reionization more
rapid in the case of more massive source halos. In particular, the
full period shown in the figures covers the redshift range of
$z=6.5-10.5$ with $V_{\rm c} = 35$ km/s but only $z=6.5-9.0$ with
$V_{\rm c} = 100$ km/s. Thus, the mass of source halos determines both
the height of the large-scale power spectrum and the redshift spread
of reionization, and independent measurements of both of these can
provide an important consistency check (though see section 4 for
additional possible complications that are not included in the model
used here).

For quantitative results, these theoretical predictions must be
compared to the expected range and sensitivity of 21-cm power spectrum
measurements. We adopt as a specific example of a first-generation
experiment the MWA, using a simple fit to the detailed sensitivity
analysis of \citet{lidz08}. Assuming their ``super-core''
configuration in which the MWA antennae are all packed close together,
we find that their results for the measurement noise $N$ in
$\Delta_{21}$ as a function of $k$ and $z$ can be fit as: \beq
\log_{10} \left( \frac{N}{2.7\ \rm mK} \right) \approx 2.8 x + 1.1 x^2
+(5 + 2.2 x) \log_{10} \left( \frac{1+z}{9} \right)\ , \eeq where $x
\equiv \log_{10}(k /(h/{\rm Mpc}))$. Over the range $z=6.8-11.5$
considered in \citet{lidz08}, this fit is accurate to within a factor
of 1.2 in $N$ for $k=0.1-0.2 h$/Mpc and a factor of 1.4 for $k=0.2-1
h$/Mpc, and corresponds to their observational parameters of a
bandwidth of 6 MHz at each redshift (i.e., a $\Delta z = 0.3$ at
$z=8$), $k$ bins of logarithmic width $d \ln k=0.5$, and 1000 hrs of
integration time (corresponding approximately to the available time
within one year of operation). We adjust the noise in $\Delta_{21}$
using its inverse square-root dependences on the bandwidth and the
$k$-bin width, and inverse dependence on the integration time
\citep{miguel05}. We use logarithmically spaced bins in both $k$ and
$1+z$, and wish to cover a broad redshift range. However, only 32 MHz
intervals of data can be computationally handled at a time by the MWA
\citep{lidz08}, so in order to avoid paying the penalty for reducing
the integration time in each frequency interval, we instead reduce the
bandwidth around each central redshift by a constant factor that makes
the sum of all the bandwidths equal to 32 MHz. We also account
approximately for the effect of foregrounds by assuming that the power
spectrum cannot be measured below $k \approx 0.1 h$/Mpc, since
foreground removal based on the smooth spectrum of the foregrounds
will also remove large-scale power in the signal.

\begin{figure}
\includegraphics[width=84mm]{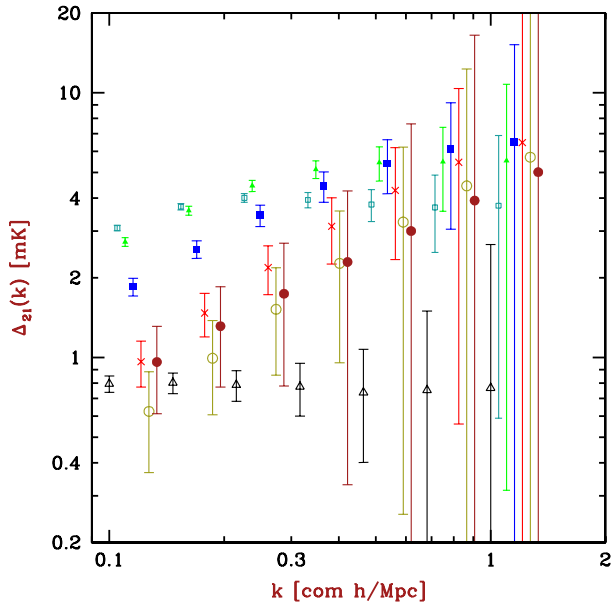}
\caption{Example of predicted values and expected observational errors
of the 21-cm power spectrum, for a year of observations with the MWA
assuming a reionization model that sets $V_{\rm c} = 35$ km/s with a
constant efficiency fixed to yield $\bar{x}^i=98\%$ at $z=6.5$. The
power spectrum is shown at 7 redshifts (out of the actual total number
of 19): $z=6.5$ ($\bar{x}^i=98\%$, open triangles), $z=7.2$
($\bar{x}^i=67\%$, open squares), $z=8.0$ ($\bar{x}^i=44\%$, closed
triangles), $z=8.9$ ($\bar{x}^i=27\%$, closed squares), $z=9.8$
($\bar{x}^i=15\%$, $\times$ symbols), $z=10.9$ ($\bar{x}^i=7.9\%$,
open circles), and $z=12$ ($\bar{x}^i=3.7\%$, closed circles). The
same 7 $k$ values are assumed at each redshift, but points at higher
$z$'s are successively offset to the right for clarity.}
\label{f:obs35}
\end{figure}

Figure~\ref{f:obs35} shows an example of the expected observational
errors in the 21-cm power spectrum, assuming a year of observations
with the MWA, at 19 central redshifts between $z=6.5$ and 12
(logarithmically spaced in $1+z$) and at 7 logarithmically spaced
central $k$ values between 0.1 and $1 h$/Mpc. The absolute errors are
large at high redshift, mainly because of the increase of the sky
temperature which is dominated by the Galactic synchrotron
emission. Towards the end of reionization, the relative errors
increase as the expected signal itself decreases. Also, the error at a
given redshift increases roughly linearly with $k$ (but faster at the
high-$k$ end) when plotted in terms of $\Delta_{21}(k)$.

\subsection{Maximum likelihood fits}

In this section we arrive at our goal of obtaining quantitative
estimates of the expected uncertainties from modeling the 21-cm power
spectrum as observed in one year of operation of the MWA. We consider
the errors both in reconstructing the reionization history and the
properties of the ionizing sources. We derive the errors from the
covariance matrix that depends on the second derivatives of $\chi^2$
near its minimum; these errors should be accurate as long as they are
small, which is the case over most of the parameter space considered
below. We consider models that specify the galactic halo population
using between 2 and 6 free parameters, as described in
section~\ref{s:model}. Unless otherwise indicated, our input models
assume constant values for $V_{\rm c}$ and $\zeta$ (corresponding to
the 2-parameter model) even when we allow additional parameters to
vary when fitting to the resulting power spectrum.

\begin{figure}
\includegraphics[width=84mm]{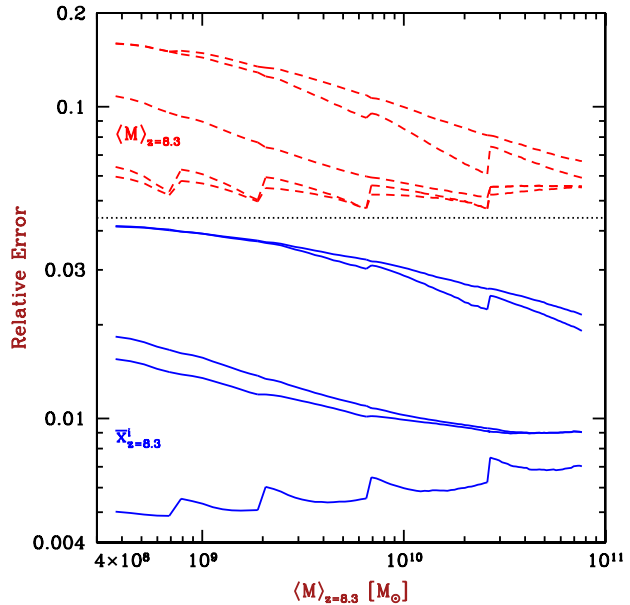}
\caption{Expected errors at the midpoint of reionization (fixed to
occur at $z=8.3$) from fitting models to 21-cm data. We consider
models with 2--6 free parameters (bottom to top in each set of
curves). A horizontal dashed line separates the two areas of the plot
that show the relative errors in the intensity-weighted mean halo mass
(top) and cosmic mean ionized fraction (bottom).}
\label{f:6-par}
\end{figure}

Figure~\ref{f:6-par} shows the expected relative errors in measuring
at $z=8.3$ the two main quantities of interest, the cosmic mean
(mass-weighted) ionization fraction $\bar{x}^i$ and the mean mass
$\langle M \rangle$ of the halos that contain the ionizing sources
(weighted by ionization intensity, which we approximate accurately as
weighting by halo mass times number density). The expected errors in
these quantities are shown as a function of $\langle M \rangle$ of the
input model, in which the assumed efficiency is set in each case to
make $\bar{x}^i = 50\%$ at $z=8.3$ (which is one of the 19 measured
redshifts). 

\begin{figure}
\includegraphics[width=84mm]{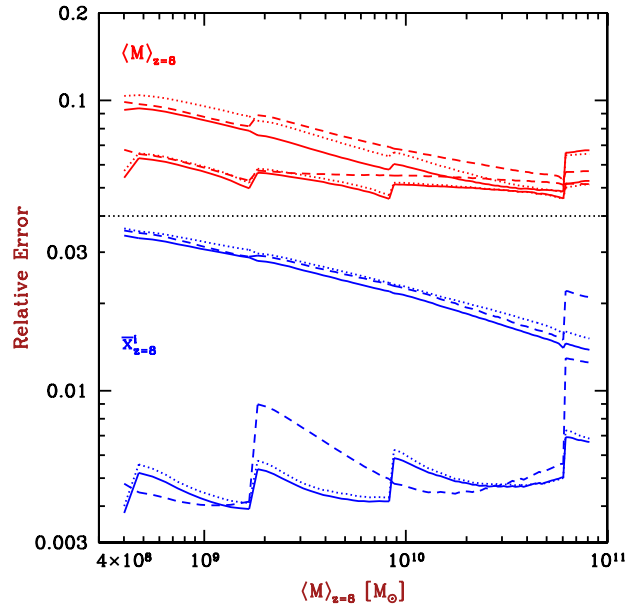}
\caption{Expected errors at $\bar{x}^i=60\%$ (fixed to occur at $z=8$)
from fitting models to 21-cm data. We consider observations at 19
redshifts and 7 wavenumbers (our standard case, solid curves), 19
redshifts and 4 wavenumbers (dotted curves), or 10 redshifts and 7
wavenumbers (dashed curves), where in all cases the assumed observing
time and the total $z$ and $k$ ranges remain the same, and the $z$ and
$k$ bins are logarithmically spaced. We only show models with either 2
or 6 parameters, where the higher curve of each pair corresponds to
having 6 parameters. A horizontal dashed line separates the two areas
of the plot that show the relative errors in $\langle M \rangle$ (top)
and $\bar{x}^i$ (bottom).}
\label{f:nbins}
\end{figure}

The halo mass range shown in Figure~\ref{f:6-par} (as well as
\ref{f:nbins}) corresponds to the range $V_{\rm c}=16.5-125$ km/s,
where the lowest value corresponds to the minimum halo mass in which
virialized gas can cool via radiative transitions in atomic hydrogen
and helium (which requires a virial temperature of $\sim 10^4$ K). For
a fixed number of parameters, the errors tend to decrease with
increasing halo mass, since if rare, massive halos drive reionization
then the halo bias is larger, increasing the large-scale ionization
fluctuations and making the power spectrum more easily observable (as
shown in the previous subsection). The curves, however, sometimes show
a jump in the error value as $\langle M \rangle$ is increased past
certain values. Each of these values corresponds to another one of the
(fixed) measured redshifts going below the end of reionization (since
the redshift $\zr$ at the end of reionization becomes closer to the
midpoint, $z=8.3$, as $\langle M \rangle$ is increased). Measurements
at $z < \zr$ do not constrain the parameters, and this reduces the
constraints on the measured parameters even at $z > \zr$; this effect
is small for the more flexible models, in which the parameters at
different redshifts are more independent of each other.

Figure~\ref{f:nbins} shows how the expected errors depend on the
number of bins used in redshift and in wavenumber. We consider a
slightly more advanced stage of reionization, $\bar{x}^i=60\%$, fixed
to occur at $z=8$ (which is one of the measured redshifts in both of
the cases that we consider here for redshift bins). For our standard
case of observations at 19 redshifts and 7 wavenumbers, the errors are
similar to those in Figure~\ref{f:6-par}, except that the error in
$\langle M \rangle$ is less dependent on the number of model
parameters, i.e., $\langle M \rangle$ is measured at $\bar{x}^i=60\%$
better than at $\bar{x}^i=50\%$ for the 6-parameter model. Reducing
the number of $k$ bins from 7 to 4 has a rather minor effect on the
errors, qualitatively consistent with \citet{lidz08} who argued that
the MWA can essentially only measure the amplitude and slope of the
21-cm power spectrum at each redshift. Reducing the number of
redshifts from 19 to 10 can have a bigger effect, producing larger
jumps in the error near some values of $\langle M \rangle$. Clearly,
it is best to divide the available 32 MHz total bandwidth into a large
number of redshift bins.  Note that we make a slight approximation in
our calculations in that we compare the data and the models at the
center of the $z$ and $k$ bins, while a more exact comparison would
average the theoretical signal over the $z$ and $k$ range within each
bin.

\begin{figure}
\includegraphics[width=84mm]{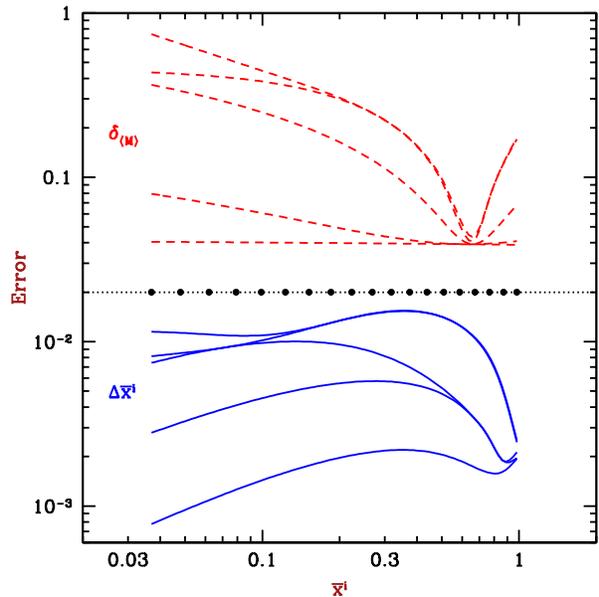}
\caption{Expected errors throughout reionization, from fitting models
to 21-cm data. We consider models with 2--6 parameters (bottom to top
in each set of curves). The input model sets $\bar{x}^i=98\%$ at
$z=6.5$ with $V_{\rm c} = 35$ km/s. A horizontal dashed line separates
the two areas of the plot that show the relative error in the
intensity-weighted mean halo mass (top) and the {\it absolute}\/ error
in the ionized fraction (bottom). Dots on this line show the values of
$\bar{x}^i$ corresponding to the 19 observed redshifts.}
\label{f:errs1}
\end{figure}

\begin{figure}
\includegraphics[width=84mm]{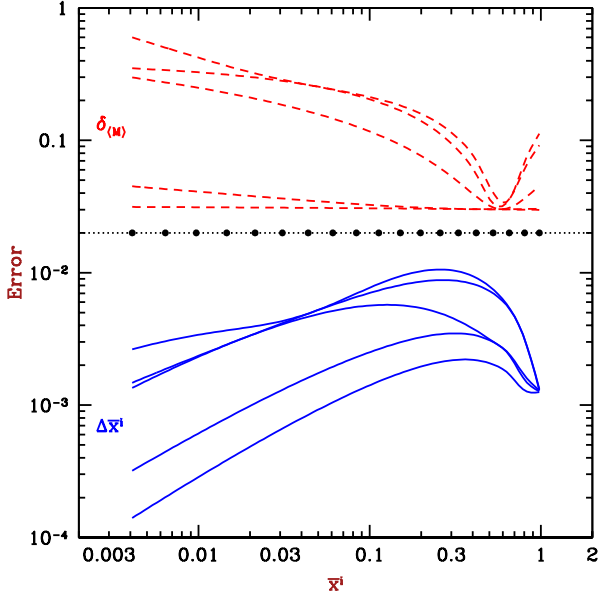}
\caption{Same as Figure~\ref{f:errs1}, except that the input model
sets $\bar{x}^i=98\%$ at $z=6.5$ with $V_{\rm c} = 100$ km/s.}
\label{f:errs2}
\end{figure}

The complete result that we can derive from our analysis is the
expected error in reconstructing the history of reionization and of
the ionizing sources within the observed redshift range of
$z=6.5-12$. This result is shown in Figures~\ref{f:errs1} and
\ref{f:errs2}, for halos with minimum $V_{\rm c} = 35$ and 100 km/s,
respectively. While we again show the relative error $\delta_{\langle
M \rangle}$, we now show the {\it absolute}\/ error $\Delta \bar{x}^i$
because this quantity varies far less during reionization than does
$\delta_{\bar{x}^i}$. The relative error $\delta_{\bar{x}^i}$
increases strongly with redshift, and becomes rather large at early
times when $\bar{x}^i$ is itself small (e.g., for $V_{\rm c} = 35$
km/s and the 6-parameter model, $\delta_{\bar{x}^i} \sim 0.3$ when
$\bar{x}^i=0.04$). Note that the same redshift range, $z=6.5$-12,
corresponds to a substantially wider portion of reionization (in terms
of the range of $\bar{x}^i$) for the more massive galactic halos.

The figures show that in general, a model with a small number of
parameters has errors that are relatively uniform throughout
reionization, since in such a restricted model the values of
$\bar{x}^i$ and $\langle M \rangle$ are strongly coupled throughout
reionization, producing similar, strongly correlated errors at all
redshifts. As the number of parameters allowed to vary freely is
increased, the expected errors obviously increase as well.

However, there are two special quantities that are well measured
regardless of the number of model parameters: the mean halo mass at
$\bar{x}^i \sim 65\%$, and $\bar{x}^i$ itself near the very end of
reionization. The measurement of $\langle M \rangle$ results mainly
from the strong dependence of the large-scale 21-cm power spectrum on
halo bias near the mid-point of reionization. The large-scale power at
this stage is the best-measured region out of all of reionization, in
terms of small relative measurement errors in $\Delta_{21}(k)$ (as
illustrated in Figure~\ref{f:obs35}). The measurement errors, which
increase rapidly with redshift due to the frequency dependence of the
galactic foreground, push the best-measured points to later times,
resulting in $\langle M \rangle$ being best measured at around the 2/3
mark of cosmic reionization. Separately, the evolution of $\bar{x}^i$
at the very end of cosmic reionization is measured accurately as well,
since the disappearance of the last remaining pockets of neutral
low-density gas causes a rapid decline with redshift in the 21-cm
power spectrum, making it highly sensitive to small changes in
$\bar{x}^i$ at a given redshift. This decline is well measured because
of the low thermal noise within the corresponding, relatively low,
redshifts (as also illustrated in Figure~\ref{f:obs35}).

We have thus far tested only input models with constant values of
$V_{\rm c}$ and $\zeta$ throughout reionization, but we now consider
also an example of a model with redshift-dependent parameters; in this
model, which is loosely motivated by feedback models, the minimum halo
$V_{\rm c}$ increases with time (perhaps due to photo-heating
feedback), decreasing the number of halos, but this is counter-acted
by an increasing ionizing efficiency with time (perhaps due to
supernova feedback being less effective in the massive halos that
dominate at later times).  Specifically, given a redshift at which
reionization is nearly complete (defined here as having
$\bar{x}^i=98\%$), we let $V_{\rm c}$ decrease linearly with redshift,
going from the atomic-cooling value of 16.5 km/s at $z=12$ to 35 km/s
when $\bar{x}^i=98\%$. Meanwhile, $\zeta$ is assumed to be
proportional to $V_{\rm c}^2$, which implies in this model a quadratic
dependence on redshift.

\begin{figure}
\includegraphics[width=84mm]{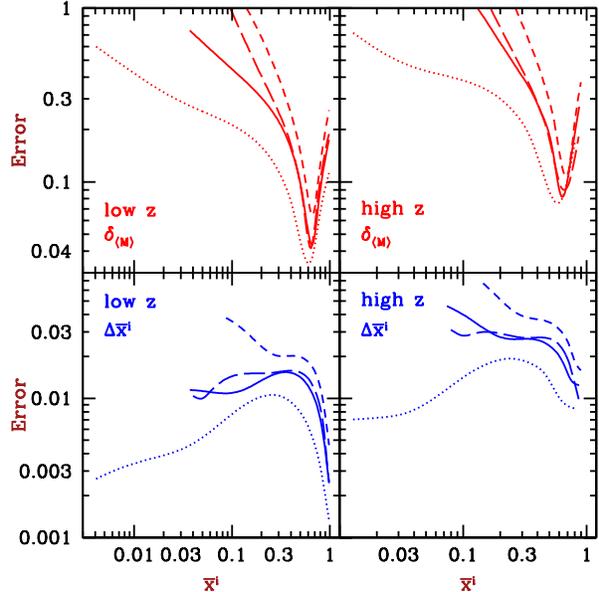}
\caption{Expected errors throughout reionization, from fitting our
full 6-parameter model to 21-cm data. We consider input models that
set $\bar{x}^i=98\%$ at redshift 6.5 (low $z$: left panels) or at
$z=7.8$ (high $z$: right panels). We consider galactic halos with
minimum $V_{\rm c} = 16.5$ km/s (short-dashed curves), 35 km/s (solid
curves), 100 km/s (dotted curves), or our feedback-inspired model
(long-dashed curves). For each model we show the relative error in the
intensity-weighted mean halo mass (top panels) and the absolute error
in the ionized fraction (bottom panels).}
\label{f:hiz}
\end{figure}

Figure~\ref{f:hiz} shows our results of fitting the 6-parameter model
to eight different input models. We consider low-mass halos
corresponding to the minimum mass for atomic cooling ($V_{\rm c} =
16.5$ km/s), intermediate-mass halos ($V_{\rm c} = 35$ km/s),
high-mass ones ($V_{\rm c} = 100$ km/s), and the feedback-inspired
model, each with reionization ending at low redshift ($\bar{x}^i=98\%$
at $z=6.5$) or at a relatively high redshift ($\bar{x}^i=98\%$ at
$z=7.8$). As already noted, the models with high-mass halos allow a
higher-precision reconstruction, especially early on in reionization
(i.e., at low $\bar{x}^i$), but with smaller errors even at the late
stages, by a factor of up to 2 though usually less, especially for
$\langle M \rangle$.  Higher-redshift reionization leads predictably
to somewhat larger errors, typically by a factor of $\sim 2$ as
expected from the increase of the sky brightness from redshift 6.5 to
7.8. The results shown at the very end of reionization are affected by
the redshift binning: while in the low-redshift models, the
lowest-redshift bin ($z=6.5$) falls right at the end of reionization
($\bar{x}^i=98\%$), the high-redshift models are an example where the
lowest-redshift bin that is still during reionization falls somewhat
earlier ($z=8$, which corresponds to $\bar{x}^i=81\%$ for $V_{\rm c} =
100$ km/s and $\bar{x}^i \sim 90\%$ for the other models).

The feedback-inspired model, which is identical to the constant
$V_{\rm c} = 35$ km/s model at the end of reionization but has lower
$V_{\rm c}$ and higher $\zeta$ values earlier on, gives errors that
are generally fairly close to those of the $V_{\rm c} = 35$ km/s
model. This indicates that there is some correlation of the
constraints at different redshifts through the model parameters, which
are limited to six (a small number compared to the total number of
measured data points); the constraints on the model parameters are
dominated by the central and late stages of reionization, where the
observational errors are small (and where the feedback-inspired model
is nearly identical to the $V_{\rm c} = 35$ km/s model).

The conclusions from these figures are rather positive in terms of the
prospects for learning about high-redshift astrophysics from the
upcoming 21-cm experiments. On the one hand, our lack of independent
knowledge of the properties of the ionizing sources has a substantial
effect on our expected ability to reconstruct both the reionization
history and the properties of the sources; in particular, allowing a
model with 6 free parameters raises the errors by up to an order of
magnitude compared to fitting with a model restricted to just 2
parameters. On the other hand, even with a 6-parameter model that
allows for a fairly large parameter space of galaxy properties, the
one-year MWA observations allow a rather impressive reconstruction of
reionization. Even the worst case in Figure~\ref{f:hiz}, of
high-redshift reionization and low-mass atomic-cooling halos, yields
relative errors of $\sim 10\%$ in $\langle M \rangle$ when $\bar{x}^i
\sim 2/3$, and $2\%$ in $\bar{x}^i$ at the end of reionization. In the
best case of low-redshift reionization (including an observed redshift
bin centered at the $98\%$ mark of cosmic reionization) together with
high-mass source halos, these relative errors drop to $4 \%$ and
$0.2\%$, respectively. The errors in $\langle M \rangle$ and
$\bar{x}^i$ are also remarkably low throughout the central stage of
reionization.

\section{Conclusions}

We have used the most realistic fully analytical model available for
the 21-cm power spectrum \citep{b07} to fit models of the galaxy
population during reionization to simulated 21-cm power spectrum
observations.  The model assumes at each redshift a fixed ionizing
efficiency $\zeta$ and a minimum halo circular velocity $V_{\rm c}$
for galactic halos.  Allowing each of these quantities to vary
linearly or quadratically with redshift yields reionization models
with up to 6 parameters, which we allow to vary freely without any
restrictions based on specific models of feedback.

Before proceeding, we compared the analytical model of \citet{b07} to
the results of three different groups that have run N-body
simulations, processed the outputs with radiative transfer, and
calculated the 21-cm power spectrum at various redshifts during the
reionization epoch. The simulations are all in reasonable agreement
with the analytical model (Figures~\ref{f:test1}--\ref{f:test3}), with
typical differences in $\Delta_{21}$ in the range $\sim 10-30\%$.
While the analytical model makes various simplifying assumptions, the
simulations are also limited by box size and radiative transfer
resolution. The analytical model generally captures the evolution of
the 21-cm power spectrum during reionization as seen in the
simulations, but a more precise comparison must await a demonstration
that the simulations have numerically converged and are consistent
with each other.

As discussed in earlier sections, the model used here should not be
considered the ultimate model to use in fitting the 21-cm power
spectrum during reionization, but as an important step towards a final
model to be constructed using guidance from numerical simulations. The
main qualitative limitation of the model used here is that while it
allows a redshift dependence in the properties of galactic halos, the
parameters are limited to being spatially uniform at each redshift. A
more realistic model would add the possibility of spatially
inhomogeneous (in particular, density-dependent) feedback. However, we
note that any such extension, which will likely add substantial
complexity to the model, must still allow a relatively quick
calculation of the 21-cm power spectrum in order to permit a maximum
likelihood analysis. It will also be important to keep the model
flexible without relying too heavily on results of particular models
or simulations, where feedback processes can only be included with
limited and approximate methods. We note that \citet{mcquinn07}
considered the effect of minihalos and Lyman-limit systems in limiting
the mean free paths of ionizing photons, and showed that their effect
on the 21-cm power spectrum at a given $\bar{x}^i$ is rather small and
is much less significant than the effect of the halo mass of the
ionizing sources.

The maximum likelihood fitting yields good grounds for optimism. While
our ignorance regarding the properties of the ionizing sources has a
substantial effect on the expected errors, we still conclude that the
expected measurements of the 21-cm power spectrum will enable us to
reconstruct both the reionization history and the properties of the
sources. In particular, even with a 6-parameter model that allows for
a fairly large parameter space of galaxy properties, the one-year MWA
observations allow a remarkably precise reconstruction of
reionization.

As a specific example of the expected errors, if reionization ends at
$z=6.5$ and is dominated by intermediate-mass halos (minimum halo
circular velocity $V_{\rm c} = 35$ km/s, corresponding to a mean halo
mass of ionizing sources $\langle M \rangle \sim 3 \times 10^9
M_\odot$ at the midpoint of reionization), then the cosmic mean
ionized fraction can be measured to $0.3\%$ accuracy at the very end
of reionization, to a relative accuracy of a few percent around the
mid-point of reionization, and better than $10 \%$ as early as a
cosmic mean ionized fraction of $\bar{x}^i = 10\%$. Also, the mean
halo mass of the ionizing sources can be measured in this case to
$5\%$ accuracy when reionization is 2/3 of the way through, and to
$20\%$ accuracy throughout the last 2/3 of reionization (i.e., when
$\bar{x}^i$ between 1/3 and 1). The errors in general increase with
the redshift at which reionization ends, and decrease with the halo
mass of the dominant ionizing sources. 

The best-measured point of reionization is around the 2/3 mark in
terms of precision in $\langle M \rangle$, and near the very end of
reionization in terms of precision in $\bar{x}^i$. The errors, though,
are fairly small in the central and late stages of reionization for
all the models that we have examined (see especially
Figure~\ref{f:hiz}), which include halos that range from the
atomic-cooling minimum ($V_{\rm c} = 16.5$ km/s, $\langle M \rangle
\sim 4 \times 10^8 M_\odot$) to 200 times more massive halos ($V_{\rm
c} = 125$ km/s, $\langle M \rangle \sim 8 \times 10^{10} M_\odot$),
examined for reionization that ends at $z \sim 6.5$ or $z \sim 8$.  We
thus conclude that if the upcoming 21-cm experiments, after
foregrounds are removed and instrumental systematics are dealt with,
reach anywhere near their expected sensitivity, then they will allow
us to study high-redshift astrophysics in unprecedented detail.


\section*{Acknowledgments}
The author is grateful for support from the ICRR in Tokyo, Japan, the
Moore Distinguished Scholar program at Caltech, and the John Simon
Guggenheim Memorial Foundation, as well as Israel Science Foundation
grant 629/05.


\label{lastpage}

\end{document}